\pdfoutput=1
\documentclass[12pt,a4paper]{article}
\usepackage{jheppub}
\usepackage[T1]{fontenc}
\usepackage{amsmath,amssymb,bm,booktabs,array,url,xspace,tikz}
\usetikzlibrary{arrows.meta,positioning,calc}
\tikzset{
  box/.style={draw, thick, rounded corners=1.5pt, minimum height=7mm, minimum width=14mm,
              inner sep=3pt, align=center},
  arr/.style={-{Latex[length=2mm]}, thick},
  lab/.style={font=\small, inner sep=1.5pt},
}
\newcommand{\SLD}{SLDiffusion\allowbreak\xspace}

\newcommand{\dd}{\mathrm{d}}
\newcommand{\tauint}{\tau_{\mathrm{int}}}
\newcommand{\Rcal}{\mathcal R}
\newcolumntype{L}[1]{>{\raggedright\arraybackslash}p{#1}}
\usepackage{xcolor}
\title{Lattice Configuration Generation with a Self-Learning Diffusion Model}
\author[a,b,c]{Akio Tomiya}
\affiliation[a]{Department of Information and Mathematical Sciences, Tokyo Woman's Christian University, Tokyo 167-8585, Japan}
\affiliation[b]{Department of Physics, Kyoto University, Kyoto 606-8502, Japan}
\affiliation[c]{RIKEN Center for Computational Science, Kobe 650-0047, Japan}
\emailAdd{akio@yukawa.kyoto-u.ac.jp}
\abstract{We show that a diffusion model for lattice field theory can be trained
by self-learning, without configurations from a separate simulation, and used
as the generator of an exact Markov chain. Training starts from exact samples
at $\beta=0$, and along a ladder of couplings the model's own chain at each
coupling, weighted by the action difference, trains the score at the next. A
Metropolis--Hastings test on each proposal and its noising path keeps the
sampled distribution exact for any network, so that network errors affect
only the acceptance and autocorrelation times. In the two-dimensional compact
XY model up to $\beta=1$, on the high-temperature side of the
Berezinskii--Kosterlitz--Thouless transition, ladders at $L=4$ and $8$
reproduce exact and reference values of the energy density and helicity
modulus at every coupling, and the generator trained from $\beta=0$ at $L=8$
has the same acceptance as one trained on an exact chain at $\beta=1$. The
acceptance follows the relation known from hybrid Monte Carlo, with terms set
by the number of noise levels and by the network error per site, and a
network trained at $L=4$ generates up to $L=16$ when the number of levels
grows with the volume.}
\begin{document}
\maketitle
\raggedbottom
\section{Introduction}
\label{sec:introduction}
Lattice field theory provides a first-principles approach to strongly
coupled quantum field theories~\cite{Wilson:1974sk,Creutz:1980zw}. When the
Boltzmann weight is nonnegative, observables are averages over field
configurations $\phi$ sampled from $p(\phi)=Z^{-1}e^{-S(\phi)}$ by
Markov chain Monte Carlo (MCMC), for which hybrid Monte Carlo
(HMC)~\cite{Duane:1987de} is the de facto standard. Near criticality,
however, the integrated autocorrelation time typically grows as a power of
the correlation length, with a dynamic critical exponent that depends on
the theory, the algorithm and the
observable~\cite{Madras:1988ei,Sokal:1997,Kennedy:2000ju}, and in lattice gauge
theories topological degrees of freedom
freeze~\cite{Schaefer:2010hu,Bonati:2017woi}. These difficulties motivate
samplers that move correlated degrees of freedom efficiently while
preserving the target distribution.

Self-learning Monte Carlo corrects the proposals of a learned effective model
with an acceptance test based on the original
action~\cite{Liu:2017slmc,Liu:2017cumulative}, and self-learning HMC retrains
its neural potential on the configurations of its own exact
chain~\cite{Nagai:2020slhmc}. We extended this approach to lattice gauge
theory with dynamical fermions, with gauge-covariant neural networks and with
equivariant Transformers~\cite{Nagai:2020jar,Nagai:2021bhh,Nagai:2023fxt}.

Normalizing
flows~\cite{Albergo:2019eim,Kanwar:2020xzo,Albergo:2021vyo,Tomiya:2025quf}
correct independent proposals with a model density computed from the
Jacobian of an invertible
map~\cite{Rezende:2020hrd,deHaan:2021erb,Singha:2022icw,R:2023dcr,Abbott:2023thq,
Komijani:2025yjz,Abbott:2025kvi,Cheng:2025khv}; related work includes
trivializing maps and their learned
versions~\cite{Luscher:2009eq,DelDebbio:2021qwf,Bacchio:2022vje,Albandea:2023wgd,Albandea:2023ais},
stochastic normalizing flows~\cite{Wu:2020snf,Caselle:2022snf}, flow-based
variance reduction~\cite{Abbott:2026ylv} and neural gauge
fixing~\cite{Hsiao:2026fdd}.

Diffusion models learn the score, which is the gradient of the logarithm
of a noise-smoothed distribution and plays the role of the force in HMC, and
generate samples by reversing the noising
process~\cite{SohlDickstein:2015,Song:2019ncsn,Ho:2020,Song:2021}. In
lattice field theory, they have been formulated as stochastic quantization
with an independence Metropolis test~\cite{Wang:2023exq}, used to generate
gauge fields by the reverse-time stochastic differential
equation~\cite{Zhu:2024kiu}, and corrected by the Metropolis-adjusted
annealed Langevin algorithm (MAALA) in the final low-noise
steps~\cite{Zhu:2025pmw,Aarts:2026zzr}. Unadjusted samplers for $SU(N)$
gauge theories~\cite{Komijani:2026lan}, force-guided group-equivariant
networks~\cite{Vega:2025hgz} and samplers for the XY model near criticality
and on large volumes~\cite{Tan:2026criticaldiffusion,Hoque:2026jnx}
followed. These samplers are trained on MCMC ensembles and learn the
distribution that the ensemble represents. Their score can be
extrapolated in the coupling~\cite{Zhu:2024kiu,Zhu:2025pmw,Aarts:2026zzr}
and transferred across
volumes~\cite{Aarts:2026zzr,Tan:2026criticaldiffusion,Hoque:2026jnx}, but
the first model needs configurations of the target distribution from a
separate simulation, and slow modes that this simulation explores poorly
can bias the training data and hence the model.

Samplers can also be trained from the unnormalized target density
alone~\cite{Richter:2024learned,Havens:2025adjoint}, for example by iterated
denoising energy matching on a replay buffer of self-generated
samples~\cite{AkhoundSadegh:2024idem}, and learned local proposals have been
built for scalar and XY lattice models~\cite{Faraz:2023xdi}. In lattice
field theory, data-free diffusion-type samplers have been built from
stochastic paths and from stochastic quantization formulated as optimal
control~\cite{Chen:2026bqw,Wang:2026pmv}. Flows have been combined with
annealed importance sampling and sequential Monte
Carlo~\cite{Arbel:2021aft,Matthews:2022sds,Midgley:2023fab} and, like
diffusion models, trained temperature by temperature on samples from the
previous temperature~\cite{Schopmans:2025tabg,Rissanen:2025ptsd,AkhoundSadegh:2025pita}.
This outer structure, a population reweighted by the change of the Boltzmann
factor and re-equilibrated by MCMC, goes back to annealed importance
sampling, population annealing and sequential Monte Carlo
samplers~\cite{Neal:2001hwj,Hukushima:2003pa,Machta:2010pa,DelMoral:2006smc}.

Proposals generated by a driven process and accepted with the ratio of path
densities go back to tempered transitions~\cite{Neal:1996tt} and
nonequilibrium candidate Monte Carlo~\cite{Nilmeier:2011ncmc}, and the path
weights of annealed importance sampling and of stochastic normalizing
flows~\cite{Wu:2020snf,Caselle:2022snf} have the same form. For diffusion
models, a reverse-diffusion proposal started from a noisy copy of the
current configuration and accepted in this way is the Metropolis-adjusted
diffusion path (MAD-Path) sampler~\cite{Chen:2026madpath}. Its ideal kernel
is a Gibbs sampler with a noisy auxiliary variable, as in diffusive Gibbs
sampling~\cite{Chen:2024digs}, and its score is trained by a variational
objective. Hill~\cite{Hill:2026ddmc} uses the same test for a global
proposal combined with local Metropolis-adjusted Langevin updates and
trains the denoiser on samples of local chains, and an extended-space
independence Metropolis--Hastings (MH) test corrects the learned forward
process of ref.~\cite{Chen:2026bqw}.

In this work we construct the self-learning diffusion sampler \SLD. It
combines reverse diffusion~\cite{Ho:2020,Song:2021}, the MH test on the
noising path~\cite{Chen:2026madpath,Hill:2026ddmc} and a score network of
effective-action form~\cite{Nagai:2021bhh,Favoni:2020reg,Nagai:2025rok}
with a new training procedure, in which the diffusion model learns on
configurations of the Markov chain that it generates. Training starts from exact samples at $\beta=0$ and
proceeds along a ladder of couplings, on which the configurations of one
coupling, weighted by the action difference, train the noise-conditioned
score at the next by denoising score matching. The network generates
proposals by reverse diffusion, from uniform noise or from a noisy copy of
the current configuration, and an MH
test on the extended space of the configuration and its noising
path accepts or rejects them. The test
uses the action and the Gaussian transition densities of the noising steps
in place of the density of the generated configuration, which reverse
diffusion does not give in closed form. The chain is therefore exact for
any network and any number of noise levels, and it supplies both the
measurements and the training data of the next coupling. The training data
thus do not come from another sampler: they start as exact samples at
$\beta=0$ and are carried along the ladder by chains that preserve the
target distribution at each coupling.

Normalizing flows trained with the reverse Kullback--Leibler
divergence~\cite{Albergo:2019eim,Kanwar:2020xzo} and the diffusion-type
sampler of ref.~\cite{Chen:2026bqw} also learn from the action alone and are
corrected by MH tests; \SLD learns from its own chain along a ladder from
$\beta=0$, without samples or a variational objective at the target
coupling. Unlike population
annealing, which estimates observables from the weighted population, \SLD
uses the weighted population only to train the generator, and unlike progressive
tempering of diffusion models~\cite{Rissanen:2025ptsd,AkhoundSadegh:2025pita},
it refreshes the training data with a chain that preserves the target
distribution.

We study the two-dimensional compact XY model up to $\beta=1$, on the
high-temperature side of its Berezinskii--Kosterlitz--Thouless (BKT)
transition at $\beta_{\rm BKT}\simeq1.12$~\cite{Hasenbusch:2005xm}, a test
bed with exact results and fast reference calculations as in the first
flow-based studies~\cite{Albergo:2019eim,Kanwar:2020xzo}. Cluster algorithms
show no visible critical slowing down for this model on lattices up to
$128^2$~\cite{Wolff:1988uh}; \SLD is aimed at theories without them, such
as gauge theories and theories with dynamical fermions.
We show that this self-learning works. Ladders from $\beta=0$ to $1$ at
$L=4$ and $8$ reproduce, at every coupling, the exact values of the energy
density and the helicity modulus from the character expansion at $L=4$ and
independent heat-bath reference values at $L=8$. At $L=8$, for proposals
from uniform noise with 256 levels, the generator of the ladder has the same
path acceptance as a network trained directly on configurations of an exact
chain at $\beta=1$, and at $L=4$ ladders repeated with different seeds give
path acceptances that agree within 0.011 at every coupling. We further show that the path acceptance
follows the relation known from HMC
(sections~\ref{sec:path-mh} and~\ref{sec:results-acceptance}), with one
term set by the number of noise levels and one by the network error per
site, which training on a larger lattice reduces. The network writes the
score as minus the gradient of a learned effective action of smeared spins
and has no parameter that refers to the lattice size. At $\beta=1$, when the
noise added to the copy has standard deviation 0.8 and the number of
levels is proportional to the volume, a network trained only at $L=4$
generates at $L=12$ and $16$ with a path acceptance of 0.41 and observables
that agree with the reference values. Finally, we combine the generator
with one sweep of a local kernel, which applies Metropolis-adjusted
Langevin steps with the learned score as drift, and find that at $L=8$,
with a network trained at that volume and a terminal noise of 3.2, this combination
gives a shorter $\tauint(\Upsilon)$ than two sweeps of the local kernel
driven by the action force (section~\ref{sec:results-production}).

This paper is organized as follows. Section~\ref{sec:xy} defines the
model and its observables, section~\ref{sec:method} the sampler,
section~\ref{sec:experiments} gives the settings of the runs, and
section~\ref{sec:results} presents the results.
Section~\ref{sec:discussion} discusses the ladder at larger volumes, the
roles of the action force and of the terminal noise, and the range of the
results, and section~\ref{sec:conclusion} concludes.
Appendix~\ref{app:exactness} discusses the ergodicity of the kernels and
the numerical evaluation of their densities, and
appendix~\ref{app:reference} the reference values.

\section{Two-dimensional compact XY model}
\label{sec:xy}
We place angular variables $\phi_x\in(-\pi,\pi]$ on an $L\times L$ square
lattice with periodic boundary conditions. With $V=L^2$, the action,
the partition function and the target distribution are
\begin{equation}
 \begin{gathered}
 S_\beta(\phi)=-\beta\sum_x\sum_{\mu=1}^2\cos(\phi_{x+\hat\mu}-\phi_x),\\
 Z_\beta=\int\mathcal D\phi\,e^{-S_\beta(\phi)},\qquad
 p_\beta(\phi)=Z_\beta^{-1}e^{-S_\beta(\phi)},
 \end{gathered}
 \label{eq:target}
\end{equation}
with the measure $\mathcal D\phi=\prod_x\dd\phi_x/(2\pi)$. The model has no
conventional long-range order at finite temperature
\cite{Mermin:1966fe,Hohenberg:1967zz}. Quasi-long-range order on the
low-temperature side and vortex unbinding on the high-temperature side
characterize the BKT transition
\cite{Berezinskii:1970pzv,Kosterlitz:1973xp,Kosterlitz:1974nba,Minnhagen:1987zz}.
In the convention of eq.~\eqref{eq:target},
$\beta_{\rm BKT}\simeq1.1199$~\cite{Hasenbusch:2005xm}.
Its compact variables and vortices, together with the uniform distribution
at $\beta=0$, from which independent configurations are drawn directly,
make the model a suitable test of self-learning.

The observables are the energy density $E$ and the helicity modulus
$\Upsilon$; the estimate $\beta_{\rm SD}$ of the coupling defined below
serves as a further check. Writing $\Delta_\mu\phi_x=\phi_{x+\hat\mu}-\phi_x$
and $\operatorname{wrap}$ for the reduction of an angular difference to
its principal interval, we define
\begin{align}
 E(\phi)=-\frac1V\sum_x\sum_{\mu=1}^2\cos(\Delta_\mu\phi_x).
 \label{eq:observables}
\end{align}
The helicity modulus~\cite{Minnhagen:1987zz},
which probes the phase stiffness, is
\begin{align}
 \Upsilon_\mu(\phi)&=\frac1V\left[\sum_x\cos(\Delta_\mu\phi_x)
 -\beta\left(\sum_x\sin(\Delta_\mu\phi_x)\right)^2\right], &
 \Upsilon&=\frac{\Upsilon_1+\Upsilon_2}{2}.
 \label{eq:helicity}
\end{align}

We test the consistency between the sampled distribution and the action
with a Schwinger--Dyson (SD) identity~\cite{Boyle:2022xor}. With
$F_x=\sum_{\mu=1}^2[\sin(\Delta_\mu\phi_x)-\sin(\Delta_\mu\phi_{x-\hat\mu})]$,
so that $\partial S_\beta/\partial\phi_x=-\beta F_x$, integration by parts over the periodic
angles gives
\begin{align}
 \Big\langle-\sum_x\frac{\partial F_x}{\partial\phi_x}\Big\rangle_\beta
 =\beta\Big\langle\sum_xF_x^2\Big\rangle_\beta,
 \label{eq:sd}
\end{align}
and the ratio $\beta_{\rm SD}$ of the sample means of the two sums
estimates the coupling at which a chain samples.

\section{Self-learning diffusion sampler}
\label{sec:method}
\SLD updates the angular field with a Markov chain whose proposals are
generated with a learned noise-conditioned score, and it trains this score
along a ladder of couplings on configurations of its own chain. We first
define the score (section~\ref{sec:score}) and the two kernels of the
chain, which leave $p_\beta$ invariant for any network
(sections~\ref{sec:path-mh} and~\ref{sec:generation-step}), and then the
ladder, the training and the network
(sections~\ref{sec:self-learning}--\ref{sec:network}).

\subsection{Noise-smoothed density and score}
\label{sec:score}
For noise of standard deviation $\sigma$, draw $z\sim\mathcal N(0,I)$ and
set
\begin{align}
 \widetilde\phi=\operatorname{wrap}(\phi+\sigma z),\qquad
 p_\sigma(\widetilde\phi\mid\phi)=\prod_xG_{\sigma^2}(\widetilde\phi_x-\phi_x),
\end{align}
where $G_v$ is the wrapped Gaussian density of variance $v$,
\begin{align}
 G_v(u)=\frac1{\sqrt{2\pi v}}\sum_{\nu\in\mathbb Z}
 \exp\left[-\frac{(u+2\pi \nu)^2}{2v}\right].
 \label{eq:wrapped}
\end{align}
The noise-smoothed target density and its score are
\begin{align}
 p_{\beta,\sigma}(\widetilde\phi)
 &=\int\mathcal D\phi\,p_\sigma(\widetilde\phi\mid\phi)\,p_\beta(\phi), &
 s^\star_{\beta,\sigma}(\widetilde\phi)
 &=\nabla_{\widetilde\phi}\log p_{\beta,\sigma}(\widetilde\phi),
 \label{eq:target-noisy-score}
\end{align}
with the measure $\mathcal D\phi$ of section~\ref{sec:xy}. As $\sigma\to0$
the score tends to the action force $-\nabla S_\beta$.
We denote by $s_\theta(\phi,\sigma;\beta)$ a network with parameters
$\theta$ that approximates $s^\star_{\beta,\sigma}$.

\subsection{Generation by reverse diffusion with path correction}
\label{sec:path-mh}
The score $s^\star_{\beta,\sigma}$ at every noise level is the quantity
needed to run a diffusion process backward~\cite{Song:2021}, so that
$s_\theta$ defines a generator of configurations. Let $0=\sigma_0<\sigma_1<\cdots<\sigma_N$ be the noise
levels of a proposal and $v_j=\sigma_j^2-\sigma_{j-1}^2$. We define the
forward and reverse transition densities on the torus by
\begin{align}
 f_j(\xi_j\mid\xi_{j-1})&=\prod_xG_{v_j}\left(\xi_{j,x}-\xi_{j-1,x}\right),
 \label{eq:forward-kernel}\\
 g_j(\xi'_{j-1}\mid\xi'_j)&=\prod_xG_{v_j}\left(\xi'_{j-1,x}-\xi'_{j,x}
 -v_js_{\theta,x}(\xi'_j,\sigma_j;\beta)\right),
 \label{eq:reverse-kernel}
\end{align}
with $G_v$ from eq.~\eqref{eq:wrapped}. From the current configuration
$\xi_0=\phi^{(t)}$ we draw a forward path $\xi_1,\ldots,\xi_N$ with
eq.~\eqref{eq:forward-kernel}. The reverse path starts from a terminal
$\xi'_N$, either drawn from the uniform distribution (\emph{uniform
terminal}) or set to $\xi'_N=\xi_N$ (\emph{shared terminal}), and
proceeds through $\xi'_{N-1},\ldots,\xi'_0$ with
eq.~\eqref{eq:reverse-kernel}. The proposal $\phi'=\xi'_0$ is accepted
with probability
\begin{align}
 P_{\rm acc}^{\rm path}=\min\left\{1,\,
 e^{-S_\beta(\xi'_0)+S_\beta(\xi_0)}\prod_{j=1}^N
 \frac{g_j(\xi_{j-1}\mid\xi_j)\,f_j(\xi'_j\mid\xi'_{j-1})}
 {f_j(\xi_j\mid\xi_{j-1})\,g_j(\xi'_{j-1}\mid\xi'_j)}\right\}.
 \label{eq:path-accept}
\end{align}
This is an MH step on the extended space of a configuration and its
forward path, whose target is $p_\beta(\xi_0)\prod_jf_j(\xi_j\mid\xi_{j-1})$.
The proposal exchanges the forward and reverse paths. Exchanging them again
inverts the ratio in eq.~\eqref{eq:path-accept}, and the density of the
uniform terminal cancels. Detailed balance therefore holds for any score
and any noise levels, and the marginal of $\xi_0$ is $p_\beta$. Errors of
the network and of the finite number of levels lower the acceptance but
leave the invariant distribution unchanged, and the positive variance of
the wrapped Gaussian transitions makes the chain irreducible and aperiodic
(appendix~\ref{app:exactness}).

The two terminals have a simple interpretation. With
$s_\theta=s^\star_{\beta,\sigma}$ and $v_j\to0$, the reverse path is
distributed as the time reversal of the forward path. For
the shared terminal, $\xi'_0$ is then a draw from
$p_\beta(\,\cdot\mid\xi_N)$ and the acceptance is one: the step is a
Gibbs update with the noisy configuration $\xi_N$ as an auxiliary
variable, and $\sigma_N$ sets the size of the move. Brownian motion on
the circle attenuates the first Fourier harmonic of an angle by
$e^{-\sigma^2/2}$, which is $0.73$ at $\sigma=0.8$ and $6\times10^{-3}$
at $\sigma=3.2$, so $\xi_N$ retains much of $\xi_0$ at $\sigma_N=0.8$ and
almost none at $\sigma_N=3.2$. For the uniform terminal, $\xi'_0$ does not
depend on $\xi_0$, and the step is an independence proposal generated by
conventional reverse diffusion. Each proposal requires the score at
$\xi'_1,\ldots,\xi'_N$ to generate the reverse path and at
$\xi_1,\ldots,\xi_N$ to evaluate $g_j$ along the forward path, that is,
$2N$ network evaluations.

One iteration of the chain draws a new forward path from the current
configuration $\phi^{(t)}$, generates $\xi'_0$, and sets
$\phi^{(t+1)}=\xi'_0$ if eq.~\eqref{eq:path-accept} accepts it and
$\phi^{(t+1)}=\phi^{(t)}$ otherwise. In some runs one sweep of the
local kernel of section~\ref{sec:generation-step} follows the path
proposal within the same iteration.

\paragraph{Acceptance and the number of levels.}
Let $\delta$ be the logarithm of the ratio in eq.~\eqref{eq:path-accept}.
It is a sum of contributions from all sites and levels, and in equilibrium
$\langle e^{\delta}\rangle=1$, as $\langle e^{-\Delta H}\rangle=1$ in HMC
for the change $\Delta H$ of the Hamiltonian along a trajectory.
If $\delta$ is Gaussian with variance $\Sigma^2$, this fixes its mean to
$-\Sigma^2/2$, and the mean acceptance is
\begin{align}
 \big\langle P_{\rm acc}^{\rm path}\big\rangle
 =\operatorname{erfc}\Big(\frac{\Sigma}{2\sqrt2}\Big),
 \qquad \Sigma^2\simeq c\,\frac{V}{N}+b\,V.
 \label{eq:acceptance-model}
\end{align}
The first relation is that of HMC,
$\langle P_{\rm acc}\rangle=\operatorname{erfc}(\tfrac12\sqrt{\langle\Delta H\rangle})$~\cite{Kennedy:2000ju},
with $\langle\Delta H\rangle=\Sigma^2/2$. Two errors contribute to
$\Sigma^2$. The finite number of levels gives a discretization error at
every site and level, which for the first-order reverse step of
eq.~\eqref{eq:reverse-kernel} adds up to a term proportional to $V/N$,
with a constant $c$ set by the spacing of the levels. The error of the
network adds a term proportional to $V$ that does not decrease with $N$,
and $b$ measures it per site. Both contributions were identified for the
same test with a variationally trained score~\cite{Chen:2026madpath}. The first term plays the role of the
step-size error of the leapfrog integrator in HMC, and the second that of
an approximate force in the molecular dynamics. As in HMC and other
Metropolis-type algorithms that update the whole lattice at once, the
acceptance at fixed parameters decreases with the volume, and increasing $N$ in
proportion to $V$ keeps the first term fixed, as decreasing the step size
does. We determine $c$ and $b$ from the measured acceptances in
section~\ref{sec:results-acceptance}, separately for the shared and the
uniform terminal, whose levels differ.

\subsection{Local kernel with the learned score}
\label{sec:generation-step}
A path proposal makes one large move, and the chain does not move when
it is rejected. The same network also supplies the drift of a local
kernel, which moves all sites by a step $h_i$ at each of $M$ noise levels
$\sigma_i$, with an MH test at each level,
\begin{align}
 \sigma_i=\sigma_1\left(\frac{\sigma_M}{\sigma_1}\right)^{(i-1)/(M-1)},
 \qquad
 h_i=\frac{A_{\rm step}\sigma_i^2}{2M},
 \qquad i=1,\ldots,M,
 \label{eq:schedule}
\end{align}
where $\sigma_1=0.05<\sigma_2<\cdots<\sigma_M=0.8$ lie within the range of
the training noise. From the current configuration $\phi$, level $i$
proposes
\begin{align}
 \phi'=\operatorname{wrap}\left(\phi+h_is_\theta(\phi,\sigma_i;\beta)
 +\sqrt{2h_i}\,\eta\right),\qquad\eta\sim\mathcal N(0,I),
 \label{eq:proposal}
\end{align}
a Langevin step whose transition density can be evaluated explicitly even
with a learned drift~\cite{Roberts:1996mala,Zhu:2025pmw},
\begin{align}
 q_i(\phi'\mid\phi)=\prod_xG_{2h_i}
 \left(\phi'_x-\phi_x-h_is_{\theta,x}(\phi,\sigma_i;\beta)\right).
 \label{eq:q}
\end{align}
The candidate is accepted with probability
\begin{align}
 P_{\rm acc}^{(i)}(\phi'\mid\phi)=\min\left\{1,
 e^{-S_\beta(\phi')+S_\beta(\phi)}\frac{q_i(\phi\mid\phi')}{q_i(\phi'\mid\phi)}\right\},
 \label{eq:accept}
\end{align}
and otherwise the configuration is kept. One sweep applies the levels in
the order $i=M,M-1,\ldots,1$, from the largest noise to the smallest,
starting from the current configuration $\phi^{(t)}$, and the
configuration after level 1 is $\phi^{(t+1)}$. Each level evaluates
the network at $\phi$ and $\phi'$, so a sweep costs $2M$ network
evaluations. Each level is an MH step with target $p_\beta$;
appendix~\ref{app:exactness}
discusses the numerical evaluation of the densities and the conditions
for convergence. We use $M=32$ and $A_{\rm step}=64$ at every coupling and
volume, which gives the proposal standard deviation
$\sqrt{2h_i}=\sqrt2\,\sigma_i$. The levels and $A_{\rm step}$ were fixed in runs
at $L=4$ and $\beta=0.3$, in which $A_{\rm step}=2$, $64$ and $128$ were
compared by the acceptance, the mean squared displacement of the angles per sweep,
$\tauint$ and $\beta_{\rm SD}$ of the chain.

This kernel is the noise-conditioned MAALA of ref.~\cite{Zhu:2025pmw} with
two changes: it starts from the current configuration of the chain
instead of a draw from a standard-normal prior, and it applies the MH test
against $p_\beta$ at every level, whereas ref.~\cite{Zhu:2025pmw} applies
it only at the final low-noise levels, after unadjusted annealed Langevin
steps through the noise-smoothed densities. The noise
levels thus vary the drift and variance of proposals that share one
invariant distribution, unlike Metropolis-adjusted samplers that use the
marginals of a diffusion process as intermediate
targets~\cite{Chen:2026madpath}.

\subsection{Self-learning with action-difference weights}
\label{sec:self-learning}
We call the sequence of couplings $0=\beta_0<\beta_1<\cdots<\beta_K$ the
\emph{ladder}. Stage $k$ trains a new network for $\beta_{k+1}$ on
configurations at $\beta_k$ (section~\ref{sec:score-training}) and then,
with this network fixed, generates an MH-corrected chain at $\beta_{k+1}$
with the kernels of sections~\ref{sec:path-mh}
and~\ref{sec:generation-step} (figure~\ref{fig:algorithm}). The training configurations of
stage $k$ form the \emph{replay}
\begin{align}
 \Rcal_k=\{\phi_k^{(n)}\}_{n=1}^{N_k}.
 \label{eq:replay}
\end{align}
The initial replay $\Rcal_0$ consists of independent configurations whose
angles are drawn from $\operatorname{Uniform}(-\pi,\pi]$. For $k\geq1$,
$\Rcal_k$ consists of states of the chain at $\beta_k$ after
thermalization, including the repetitions after rejected proposals.

Configurations at $\beta_k$ represent $p_{\beta_{k+1}}$ once they are
weighted with
\begin{align}
 w_k(\phi)=\exp[-S_{\beta_{k+1}}(\phi)+S_{\beta_k}(\phi)],
 \label{eq:weights}
\end{align}
since, for any integrable observable
$O$~\cite{Ferrenberg:1988reweighting,Iwami:2015multipoint},
\begin{align}
 \langle O\rangle_{\beta_{k+1}}
 =\frac{\langle w_k O\rangle_{\beta_k}}{\langle w_k\rangle_{\beta_k}}.
 \label{eq:reweighting}
\end{align}
On the finite replay we approximate eq.~\eqref{eq:reweighting} with the
normalized weights
\begin{align}
 \widehat w_{k,n}=\frac{w_k(\phi_k^{(n)})}{\sum_mw_k(\phi_k^{(m)})}.
 \label{eq:normalized-weights}
\end{align}
We use them without clipping to train the score at $\beta_{k+1}$. The
weight $w_k$ is the incremental weight of annealed importance sampling and
population annealing~\cite{Neal:2001hwj,Hukushima:2003pa,Machta:2010pa}.
Observables are measured without weights on the MH-corrected chain at
$\beta_{k+1}$.

We measure the concentration of the weights by their effective sample
size
\begin{align}
 N_w=\Big(\sum_n\widehat w_{k,n}^2\Big)^{-1}.
 \label{eq:weight-ess}
\end{align}
It measures only the spread of the weights; the autocorrelation of the
chain lowers the number of independent configurations in the replay
further. Stage $k$ is trained only if $N_w\geq2000$ and $N_w/N_k\geq0.10$.
Otherwise the step from $\beta_k$ to $\beta_{k+1}$ is too large for the
replay, and we take a smaller step from the same chain
(section~\ref{sec:experiments}). The chain at $\beta_{k+1}$ becomes the
next replay without further tests.

\begin{figure}[t]
\centering
\begin{tikzpicture}[node distance=9mm and 9mm]
 \node[box] (source) {replay $\Rcal_k$};
 \node[box, right=of source] (weight) {$\widehat w_{k,n}\propto w_k(\phi_k^{(n)})$};
 \node[box, right=of weight] (learn) {train $s_\theta(\phi,\sigma;\beta_{k+1})$};
 \node[box, below=of learn] (gen) {path proposals at $\beta_{k+1}$};
 \node[box, below=of weight] (chain) {chain $\phi^{(t)}$};
 \node[lab, above=of source] (init) {$\Rcal_0$: uniform, $\beta_0=0$};
 \draw[arr] (init) -- (source);
 \draw[arr] (source) -- (weight);
 \draw[arr] (weight) -- (learn);
 \draw[arr] (learn) -- node[lab, right] {$\theta^*_{k+1}$} (gen);
 \draw[arr] (gen) -- (chain);
 \draw[arr] (chain) -| node[lab, pos=0.75, left] {$k\to k+1$} (source);
 \draw[arr] (chain) -- ++(0,-12mm) node[below] {$\langle O\rangle_{\beta_{k+1}}$};
\end{tikzpicture}
\caption{One stage of the self-learning ladder. The replay $\Rcal_k$ of
eq.~\eqref{eq:replay}, initially independent uniform configurations at
$\beta_0=0$, is weighted with eq.~\eqref{eq:weights} to train the score at
$\beta_{k+1}$. With the trained network fixed, path proposals
(section~\ref{sec:path-mh}), optionally followed by one sweep of the local
kernel per iteration (section~\ref{sec:generation-step}), generate an MH-corrected chain
at $\beta_{k+1}$. Its configurations are the next replay, and observables
are measured on the same chain without weights.}
\label{fig:algorithm}
\end{figure}

\subsection{Weighted training of the score}
\label{sec:score-training}
On the weighted replay the smoothed density is
\begin{align}
 \widehat p_{k+1,\sigma}(\widetilde\phi)
 &=\sum_n\widehat w_{k,n}\,p_\sigma(\widetilde\phi\mid\phi_k^{(n)}),
 \label{eq:weighted-noisy-density}
\end{align}
and its score, which we call the \emph{reweighted score}, is
\begin{align}
 \widehat s^{\rm rw}_{k+1,\sigma}(\widetilde\phi)
 &=\nabla_{\widetilde\phi}\log\widehat p_{k+1,\sigma}(\widetilde\phi)
 =\sum_n\omega_{k,n}(\widetilde\phi,\sigma)\,
 s_\sigma(\widetilde\phi\mid\phi_k^{(n)}),
 \label{eq:reweighted-score}
\end{align}
with the posterior weights and the score of the noise kernel
\begin{align}
 \omega_{k,n}(\widetilde\phi,\sigma)
 &=\frac{\widehat w_{k,n}p_\sigma(\widetilde\phi\mid\phi_k^{(n)})}
 {\sum_m\widehat w_{k,m}p_\sigma(\widetilde\phi\mid\phi_k^{(m)})}, &
 s_\sigma(\widetilde\phi\mid\phi)
 &=\nabla_{\widetilde\phi}\log p_\sigma(\widetilde\phi\mid\phi).
 \label{eq:teacher-score}
\end{align}

The global shift $\phi_x\to\phi_x+\alpha$ leaves $p_\beta$
invariant, so the components of $s^\star_{\beta,\sigma}$ sum to zero. The
reweighted score of the finite replay can have a uniform component,
which we remove with the projection
\begin{align}
 (\Pi_0s)_x=s_x-\frac1V\sum_ys_y.
 \label{eq:zero-mode-projector}
\end{align}
The network of section~\ref{sec:network} has zero component sum by
construction. We train it with the denoising score-matching
objective~\cite{Vincent:2011dsm,Song:2021}, which avoids evaluating the
replay sum of eq.~\eqref{eq:reweighted-score} at every training step,
\begin{align}
 \widehat{\mathcal L}_{k+1}(\theta)
 =\sum_n\widehat w_{k,n}\,\mathbb E_{\sigma,z}
 \left[\frac1V\left\|
 \sigma s_\theta(\widetilde\phi,\sigma;\beta_{k+1})
 -\sigma\Pi_0s_\sigma(\widetilde\phi\mid\phi_k^{(n)})
 \right\|^2\right].
 \label{eq:weighted-dsm}
\end{align}
Over zero-sum vector fields its minimizer is
$\Pi_0\widehat s^{\rm rw}_{k+1,\sigma}$, the conditional mean of the
projected score of the noise kernel under the weighted joint distribution
of $\phi_k^{(n)}$ and $\widetilde\phi$.

\paragraph{Training target.}
At small $\sigma$ the target $\sigma\Pi_0s_\sigma$ is dominated by the
noise $z$, whose variance per component is one, while the quantity it
estimates, $\sigma$ times the score, vanishes as $\sigma\to0$. Since the
action is known, we replace the target in eq.~\eqref{eq:weighted-dsm} by
\begin{align}
 \lambda_\sigma\,\Pi_0\big(-\nabla S_{\beta_{k+1}}(\phi_k^{(n)})\big)
 +(1-\lambda_\sigma)\,\Pi_0s_\sigma(\widetilde\phi\mid\phi_k^{(n)}),
 \qquad \lambda_\sigma=\frac{1}{1+2\beta_{k+1}^2\sigma^2}.
 \label{eq:tsm-blend}
\end{align}
For a translation-invariant noise kernel on the torus, integration by
parts gives $\mathbb E[-\nabla S_\beta(\phi)\mid\widetilde\phi]
=\nabla\log p_{\beta,\sigma}(\widetilde\phi)$, the identity behind Target
Score Matching~\cite{DeBortoli:2024tsm}. Both terms of
eq.~\eqref{eq:tsm-blend} therefore have the same conditional mean, and
the blend leaves the minimizer unchanged. The first term has small
variance at small $\sigma$ and the second at large $\sigma$; the weight
$\lambda_\sigma$ combines them in inverse proportion to their variances
per component, estimated as $1$ for $\sigma s_\sigma$ and
$2\beta^2\sigma^2$ for $\sigma\nabla S_\beta$. Control Variate Score
Matching~\cite{Kahouli:2025cvsm} combines the two identities in a related
way. The force in eq.~\eqref{eq:tsm-blend} is evaluated at $\beta_{k+1}$
on replay configurations that enter with the weights $\widehat w_{k,n}$,
so that they represent $p_{\beta_{k+1}}$, the distribution for which
the identity holds.

\paragraph{One training step.}
A minibatch is formed by drawing replay indices with replacement in
proportion to $\widehat w_{k,n}$, so that its mean squared error
estimates eq.~\eqref{eq:weighted-dsm}. For each configuration, $\sigma$
is drawn log-uniformly from $[0.05,3.2]$, the configuration is transformed
by a random rotation or reflection of the square lattice and a random sign
$\phi\to-\phi$, and a fresh $z$ gives
$\widetilde\phi$ and $\sigma$ times the target of
eq.~\eqref{eq:tsm-blend}. One training step is one AdamW update of $\theta$
on the minibatch mean of the squared deviation of $\sigma s_\theta$ from
this target. We keep an exponential moving average of the parameters and
select the averaged parameters $\theta^*_{k+1}$ with the lowest loss on
held-out configurations at $\beta_k$, weighted with
eq.~\eqref{eq:weights}. The optimizer settings are given in
section~\ref{sec:experiments}. The path proposals and the local kernel at
$\beta_{k+1}$ use $s_\theta$ with the parameters $\theta^*_{k+1}$.

\subsection{Score network from a smeared effective action}
\label{sec:network}
The smoothed density in eq.~\eqref{eq:target-noisy-score} solves the heat
equation $\partial p_{\beta,\sigma}/\partial(\sigma^2/2)=\nabla^2p_{\beta,\sigma}$
with the initial condition $p_\beta$. For small $\sigma$, and up to
corrections from the periodic images that are exponentially small in
$1/\sigma^2$,
\begin{align}
 \log p_{\beta,\sigma}(\phi)
 =-S_\beta(\phi)
 +\frac{\sigma^2}{2}\Big(\big|\nabla S_\beta(\phi)\big|^2-\nabla^2S_\beta(\phi)\Big)
 +O(\sigma^4)+\text{const}.
 \label{eq:small-sigma}
\end{align}
For the action in eq.~\eqref{eq:target}, both corrections are sums of
local terms: $|\nabla S_\beta|^2$ is the sum of the squared forces on the
sites, and $\nabla^2S_\beta=-2S_\beta(\phi)$. Higher orders bring in larger
but still finite neighbourhoods. This motivates writing the score at every
$\sigma$, like $s^\star_{\beta,\sigma}$, as minus the gradient of an
effective action that is a sum of local, symmetric terms,
\begin{align}
 s_\theta(\phi,\sigma;\beta)=-\nabla_\phi S^{\rm eff}_\theta(\phi;\sigma,\beta),\qquad
 S^{\rm eff}_\theta(\phi;\sigma,\beta)=\sum_x\ell_\theta\big(\psi_x(\Phi(\phi))\big),
 \label{eq:effective-action-score}
\end{align}
where $\Phi$ is a learned smearing of the angles and $\psi_x$ are local
invariants of the smeared field. Since $S^{\rm eff}_\theta$ is invariant
under the global shift $\phi_x\to\phi_x+\alpha$,
$\Pi_0s_\theta=s_\theta$ with $\Pi_0$ of eq.~\eqref{eq:zero-mode-projector}.

The smearing follows the covariant networks for lattice gauge fields
\cite{Nagai:2021bhh,Favoni:2020reg} and their attention-based extension,
the gauge-covariant transformer CASK~\cite{Nagai:2025rok}. In the XY
model the spin $e^{i\phi_x}$ plays the role of the link variable, and
the spins at the offsets $d\in\mathcal D$ play the role of the staples. We
use three shells of offsets: nearest neighbours ($\pm\hat\mu$), diagonal
neighbours ($\pm\hat1\pm\hat2$), and second neighbours along the axes
($\pm2\hat\mu$), twelve offsets in total. One smearing layer reads
\begin{align}
 \phi^{(l+1)}_x&=\phi^{(l)}_x
 +\sum_{d\in\mathcal D}a^{(l)}_{x,d}\,\gamma^{(l)}_{r(d)}
 \sin\big(\phi^{(l)}_{x+d}-\phi^{(l)}_x\big), \\
 a^{(l)}_{x,d}&=\frac{\exp\big[\kappa^{(l)}\cos(\phi^{(l)}_{x+d}-\phi^{(l)}_x)+\zeta^{(l)}_{r(d)}\big]}
 {\sum_{d'\in\mathcal D}\exp\big[\kappa^{(l)}\cos(\phi^{(l)}_{x+d'}-\phi^{(l)}_x)+\zeta^{(l)}_{r(d')}\big]},
 \label{eq:smearing}
\end{align}
where $r(d)\in\{1,2,3\}$ is the shell of the offset $d$. The first line is
a stout step~\cite{Morningstar:2003gk}, and the weights $a^{(l)}_{x,d}$
are an attention over the neighbours that depends on how well each
neighbour is aligned. After $n_{\rm L}$ layers,
$\Phi(\phi)=\phi^{(n_{\rm L})}$. At each site we form six invariants of
$\Phi$: the shell sums $e_{r,x}=\frac14\sum_{d:\,r(d)=r}\cos(\Phi_{x+d}-\Phi_x)$
for $r=1,2,3$, and $F_{1,x}^2/4$, $F_{2,x}^2/4$ and $F_{1,x}F_{2,x}/4$ with
$F_{r,x}=\sum_{d:\,r(d)=r}\sin(\Phi_{x+d}-\Phi_x)$. On the unsmeared field,
$F_{1,x}$ is the $F_x$ of section~\ref{sec:xy} and $\beta F_{1,x}$ is the
force on site $x$, so the leading terms of eq.~\eqref{eq:small-sigma} are
linear combinations of $e_{1,x}$ and $F_{1,x}^2$. The head $\ell_\theta$ is a
linear function of $\psi_x$ plus a two-layer perceptron with feature-wise
linear modulation (FiLM)~\cite{Perez:2018film}.

All coefficients ($\gamma^{(l)}_r$, $\kappa^{(l)}$, $\zeta^{(l)}_r$, the
linear coefficients and the FiLM scale and shift of the head) are outputs
of a small perceptron of $(\log\sigma,\beta)$ with random Fourier
features~\cite{Tancik:2020fourier}. Since the coefficients depend only on
$\sigma$, $\beta$ and the shell, and the invariants are built from shell
sums and are even under $\phi\to-\phi$, the network has the
symmetries of the target: the global $U(1)$ shift, lattice translations,
the rotations and reflections of the square lattice, and
$\phi\to-\phi$, under which the score changes sign. It contains no
parameter that refers to the lattice size, so a network trained at one
$L$ can be evaluated at any other
(section~\ref{sec:results-network}). We compute $\nabla_\phi S^{\rm eff}_\theta$
by the chain rule through the head and the smearing layers, in the same
way as the force of a smeared action in HMC, so that automatic
differentiation is needed only once, with respect to $\theta$. With
$n_{\rm L}=3$ smearing layers and hidden width 32 the network has 4{,}923
parameters. One evaluation at $L=8$ takes about $0.1$\,ms on a single CPU
core.

\section{Simulation setup}
\label{sec:experiments}
\subsection{Ladders and training}
We run the ladder of section~\ref{sec:self-learning} at $L=4$ and $L=8$
(table~\ref{tab:settings}). At $L=4$ the couplings are
\begin{align}
 0\longrightarrow0.15\longrightarrow0.3\longrightarrow0.45\longrightarrow0.6
 \longrightarrow0.75\longrightarrow0.9\longrightarrow1.
 \label{eq:beta-grid}
\end{align}
At $L=8$ the couplings are the same up to $\beta=0.6$. Above $\beta=0.6$,
where the step to $0.75$ gives $N_w/N_k$ between 0.061 and 0.071 in the
three ladders of section~\ref{sec:results-L8}, below the threshold of
section~\ref{sec:self-learning}, the next coupling is the largest value on
a grid of spacing 0.025 for which the weights on the current chain give
$N_w/N_k\geq0.25$, a margin above that threshold; this gives $0.7$, $0.8$, $0.9$ and $1$. The first stage
trains on independent uniform configurations, and every later stage on
the configurations of the chain at the previous coupling. The first 90\%
of them, $N_k=18{,}000$, form the replay,
and the last 10\% are held out for validation; both enter with the weights
of eq.~\eqref{eq:weights}.

\begin{table}[t]
\centering\small
\begin{tabular}{L{38mm} L{92mm}}
\toprule
Ladders & $L=4$: eq.~\eqref{eq:beta-grid}; $L=8$: steps of 0.15 to $0.6$, then $0.7$, $0.8$, $0.9$, $1$\\
Training per stage & 20{,}000 configurations, 90\% as replay; 20{,}000 steps\\
Path proposal, ladders & uniform terminal; $N$ levels geometric in $[0.05,3.2]$; $N=64$ at $L=4$, $256$ at $L=8$\\
Path proposal, $\beta=1$ & shared terminal; $N$ levels geometric in $[0.05,0.8]$; $N=64$, $128$, $256$ at $L=8$, $12$, $16$\\
Local sweep & $M=32$; $\sigma_i$ geometric from $0.05$ to $0.8$, applied from $0.8$ down; $A_{\rm step}=64$\\
Thermalization & 1{,}000 iterations from a uniform configuration\\
Measurement & every iteration; 20{,}000 iterations per chain; for the checks at $L=4$, $6\times250{,}000$ at each of $\beta=0.3$ and $1$\\
\bottomrule
\end{tabular}
\caption{Settings of the ladders and of the production chains in
section~\ref{sec:results}. The chains of
section~\ref{sec:results-acceptance} and the chain with $\sigma_N=3.2$ in
section~\ref{sec:results-production} vary $N$ and $\sigma_N$ as stated
there. An iteration is defined in section~\ref{sec:path-mh}.}
\label{tab:settings}
\end{table}

Every stage trains a new network from random initialization, with the
architecture of section~\ref{sec:network}, $n_{\rm L}=3$ smearing layers
and width 32; at $L=8$ a third ladder uses a larger network with
$n_{\rm L}=5$ and width 64 (10{,}857 parameters). Training uses
minibatches of 64, AdamW~\cite{Loshchilov:2019adamw}
with learning rate $2\times10^{-3}$ and weight decay $10^{-4}$, gradient
clipping at norm one, a linear warm-up over the first 5\% of the steps
followed by a cosine decay to 5\% of the initial rate, and parameter
averaging with decay 0.999. Every 800 steps the averaged parameters are
evaluated on 256 validation configurations, each noised at the 13 levels
$\sigma\in\{0.05,0.08,0.12,0.18,0.27,0.40,0.60,0.80,1.10,1.50,2.00,2.60,3.20\}$,
and the parameters with the lowest validation loss are kept. One stage of
training took about 6 minutes at $L=4$ and 16 to 20 minutes at $L=8$ (27
to 30 minutes for the larger network) on a single CPU core.

For sections~\ref{sec:results-L8}--\ref{sec:results-acceptance} we also
trained networks with $n_{\rm L}=3$ directly at $\beta=1$, with the same
settings, on the production chains of
section~\ref{sec:results-production}: at $L=8$ on
the 20{,}000 configurations of one chain, and at $L=16$ on the 60{,}000
configurations of three chains with 40{,}000 steps. We call them the
networks trained at $L=8$ and at $L=16$. Two variants of the network
trained at $L=8$ use 60{,}000 steps, one on the same configurations and
one on the 80{,}000 configurations of the four chains with the local
sweep at $L=8$. Score errors of different networks are compared on
configurations of production chains on which none of them was trained.

\subsection{Generation settings}
The path acceptance is the fraction of accepted path proposals. In the
runs labelled ``+\,local'', each iteration (section~\ref{sec:path-mh}) adds
one sweep of the local kernel of section~\ref{sec:generation-step} with the
score of the same network. The production chains at $\beta=1$ on $L=8$,
$12$ and $16$ (section~\ref{sec:results-production}) use, without
retraining, the last-stage network of the $L=4$ ladder that was run with
the generator alone; we call it the network of the $L=4$ ladder. As a
control, the local
kernel runs alone, one sweep per iteration, with the action force
$-\nabla S_\beta$, the limit of the score as $\sigma\to0$
(section~\ref{sec:score}), as drift and no network. The image sums of the path
proposals are truncated at $|\nu|\leq5$, and every run checks that two
further images change the logarithm of the density by less than
$10^{-12}$; the local kernel uses $|\nu|\leq3$
(appendix~\ref{app:exactness}).

The numbers of levels and the terminal noise were fixed in test runs.
The ladders use $N=64$ at $L=4$, which gives a path acceptance of 0.37 at
$\beta=1$ (section~\ref{sec:results-L4}), and
$N=256$ at $L=8$, which keeps $V/N=1/4$. With the shared terminal we use
$\sigma_N=0.8$ and choose $N$ so that $V/N\approx1$.
Section~\ref{sec:results-acceptance} shows how the acceptance depends on
$N$ and on the terminal.

Training, the choice of couplings and checkpoints, and the generation
settings use only the action and configurations of self-learning chains.
Exact values at $L=4$ from the character expansion and an independent
heat-bath and overrelaxation calculation at $L=8$, $12$ and $16$
(appendix~\ref{app:reference}) serve only to validate the results.
All runs use Julia~1.11.9~\cite{Bezanson:2014pyv}, with the packages
Zygote.jl for the gradients with respect to $\theta$ and Optimisers.jl for
AdamW, and one thread each on an Apple M3 Ultra.

\subsection{Autocorrelation and statistical errors}
\label{sec:errors}
Following refs.~\cite{Madras:1988ei,Sokal:1997,Luscher:2004pav}, we
define for measurements $O_1,\ldots,O_T$ of an observable $O$ the
autocorrelation function and the integrated autocorrelation time,
\begin{align}
 \Gamma_O(t)&=\frac1{T-t}\sum_{n=1}^{T-t}(O_n-\overline O)(O_{n+t}-\overline O),
 \label{eq:acf}\\
 \tauint(O)&=\frac12+\sum_{t=1}^W\frac{\Gamma_O(t)}{\Gamma_O(0)}.
 \label{eq:tau}
\end{align}
With this normalization independent measurements have $\tauint=1/2$, and
the variance of the mean is $2\tauint\Gamma_O(0)/T$ for $T\gg\tauint$.
The window $W$ is the smallest of three lags: the smallest lag that
satisfies the self-consistent condition $W\geq5\,\tauint(W)$ of
ref.~\cite{Sokal:1997}, the last lag before $\Gamma_O(t)$ first becomes
non-positive, and 5{,}000. Autocorrelation times are given in units of
iterations.

For a single chain we quote the error of the mean from blocks of 500
consecutive iterations, more than 100 times $\tauint$ of these chains.
Where several chains with different seeds are available, we quote their
mean, the standard error from the spread between chains, and the mean of
$\tauint$ over the chains. These errors are statistical at fixed networks; section~\ref{sec:results-L4}
gives the variation under retraining.

The cost of an iteration, $2N$ network evaluations for a path proposal
and $2M=64$ for a local sweep (sections~\ref{sec:path-mh}
and~\ref{sec:generation-step}), differs between kernels. The ratio of
autocorrelation times per iteration therefore compares the mixing of
kernels on the same lattice and is not a measure of relative
efficiency~\cite{Albergo:2019eim}; for samplers that include training,
the cost also depends on the training protocol~\cite{Abbott:2022zsh}.

\section{Numerical results}
\label{sec:results}

\subsection{Score networks across volumes}
\label{sec:results-network}
Table~\ref{tab:network-transfer} compares the network of
section~\ref{sec:network} with a convolutional network of 690{,}194
parameters. Its input channels are $\sin\Delta_\mu\phi_x$,
$\cos\Delta_\mu\phi_x$ and local functions of the angles, a periodic
convolution is followed by eight residual blocks with group
normalization~\cite{Wu:2018groupnorm} and FiLM conditioning on $\sigma$,
and the output is the lattice divergence of link currents, so that the
components of the score sum to zero. Both networks are trained at
$L=4$ and $\beta=1$ on the configurations of one chain at that coupling,
with the same loss, noise distribution, random symmetry transformations
of the configurations, optimizer and number of training steps. They are
then evaluated without retraining at $L=4$, $8$ and $12$ on configurations
of other chains. The entry is the mean squared deviation of $\sigma
s_\theta$ from the training target per component. It contains the variance
of the target, which no network can remove. At $L=4$ all networks
reach the same error. At $L=8$ and $12$ the error of the convolutional
network grows by a factor of 5 to 17 for $\sigma\leq0.27$ and of 3 at
$\sigma=0.8$, while that of the effective-action network grows by at most
about 25\%. The convolutional network with per-site normalization, which
has no operation over the whole lattice, degrades in the same way as the
one with group normalization. This suggests that the difference comes
from the form of eq.~\eqref{eq:effective-action-score}, which restricts
the score to the gradient of a sum of local terms.

\begin{table}[t]
\centering\small
\setlength{\tabcolsep}{5pt}
\begin{tabular}{llcccccc}\toprule
network & $L$ & $\sigma=0.05$ & $0.12$ & $0.27$ & $0.8$ & $1.5$ & mean\\\midrule
convolutional, group norm & 4 & 0.0002 & 0.0007 & 0.0112 & 0.122 & 0.340 & 0.0971\\
 & 8 & 0.0015 & 0.0116 & 0.0641 & 0.388 & 0.487 & 0.1792\\
 & 12 & 0.0013 & 0.0104 & 0.0584 & 0.376 & 0.488 & 0.1749\\\midrule
convolutional, per-site norm & 4 & 0.0003 & 0.0007 & 0.0112 & 0.121 & 0.342 & 0.0972\\
 & 8 & 0.0018 & 0.0128 & 0.0688 & 0.410 & 0.503 & 0.1869\\
 & 12 & 0.0014 & 0.0116 & 0.0635 & 0.395 & 0.506 & 0.1819\\\midrule
effective action, $n_{\rm L}=0$ & 4 & 0.0001 & 0.0007 & 0.0111 & 0.121 & 0.337 & 0.0970\\
 (2{,}262 parameters) & 8 & 0.0001 & 0.0007 & 0.0126 & 0.150 & 0.372 & 0.1046\\
 & 12 & 0.0001 & 0.0008 & 0.0120 & 0.153 & 0.381 & 0.1061\\\midrule
effective action, $n_{\rm L}=3$ & 4 & 0.0002 & 0.0007 & 0.0112 & 0.119 & 0.336 & 0.0964\\
 (4{,}923 parameters) & 8 & 0.0002 & 0.0007 & 0.0126 & 0.145 & 0.357 & 0.1016\\
 & 12 & 0.0002 & 0.0008 & 0.0120 & 0.147 & 0.367 & 0.1033\\
\bottomrule\end{tabular}
\caption{Validation error of networks trained at $L=4$, $\beta=1$ and
evaluated at $L=4$, $8$, $12$ without retraining. Entries are the mean
squared deviation of $\sigma s_\theta$ from the training target per
component, on 256 held-out configurations at each $L$, for five of the 13 validation noise levels of
section~\ref{sec:experiments}; ``mean'' averages over all 13.
The convolutional networks have 690{,}194 parameters, and the
per-site normalization replaces each group normalization by the division
of the channel vector at each site by its root mean square.}
\label{tab:network-transfer}
\end{table}

\subsection{Ladders at \texorpdfstring{$L=4$}{L=4}}
\label{sec:results-L4}
At $L=4$ the energy density and the helicity modulus are known exactly at
every coupling (appendix~\ref{app:reference}), so every stage of the
ladder can be checked. We ran the ladder of eq.~\eqref{eq:beta-grid}
with the generator alone and with the local sweep, and repeated the ladder
with the generator alone twice, with the same settings and different
seeds, to measure the variation under retraining;
table~\ref{tab:ladder-L4} shows the first of the three ladders with the
generator alone. The relative
effective sample size $N_w/N_k$ of eq.~\eqref{eq:weight-ess} lies between
0.49 and 0.87 at all stages. The two ladders of table~\ref{tab:ladder-L4}
reproduce the exact energy
density and helicity modulus at all seven couplings
(table~\ref{tab:ladder-L4}). For seven degrees of freedom, $\chi^2$ is
$9.1$ for the energy and $10.3$ for the helicity modulus with the
generator alone, and $5.6$ and $5.6$ with the local sweep; the largest
deviation is 2.0 standard errors. With the generator alone, whose
proposals do not depend on the current configuration, $\tauint(E)$ grows
from 0.6 to 2.1 iterations as the path acceptance falls; one local sweep
per iteration keeps it below 0.75.

\begin{table}[t]
\centering\small
\begin{tabular}{l c ccc c ccc}
\toprule
 & & \multicolumn{3}{c}{generator} & & \multicolumn{3}{c}{generator + local}\\
\cmidrule{3-5}\cmidrule{7-9}
$\beta$ & $E_{\rm exact}$ & acc. & $\langle E\rangle$ & $\tauint(E)$ & & acc. & $\langle E\rangle$ & $\tauint(E)$\\
\midrule
0.15 & $-0.15218$ & 0.912 & $-0.1538(18)$ & 0.62 & & 0.913 & $-0.1495(20)$ & 0.51\\
0.3 & $-0.31933$ & 0.824 & $-0.3153(24)$ & 0.75 & & 0.819 & $-0.3212(16)$ & 0.53\\
0.45 & $-0.52512$ & 0.723 & $-0.5234(29)$ & 0.97 & & 0.715 & $-0.5254(23)$ & 0.59\\
0.6 & $-0.79335$ & 0.602 & $-0.7907(32)$ & 1.29 & & 0.599 & $-0.7912(24)$ & 0.68\\
0.75 & $-1.08501$ & 0.488 & $-1.0784(44)$ & 1.50 & & 0.485 & $-1.0870(28)$ & 0.72\\
0.9 & $-1.30973$ & 0.405 & $-1.3115(40)$ & 1.98 & & 0.412 & $-1.3076(22)$ & 0.72\\
1 & $-1.41173$ & 0.371 & $-1.4066(35)$ & 2.09 & & 0.373 & $-1.4114(22)$ & 0.66\\
\midrule
$\beta$ & $\Upsilon_{\rm exact}$ & & $\langle\Upsilon\rangle$ & $\tauint(\Upsilon)$ & & & $\langle\Upsilon\rangle$ & $\tauint(\Upsilon)$\\
\midrule
0.15 & 0.00179 & & 0.0012(12) & 0.61 & & & 0.0001(12) & 0.53\\
0.3 & 0.01687 & & 0.0137(20) & 0.75 & & & 0.0161(15) & 0.52\\
0.45 & 0.07238 & & 0.0694(30) & 0.93 & & & 0.0719(21) & 0.56\\
0.6 & 0.20898 & & 0.2057(36) & 1.21 & & & 0.2052(25) & 0.64\\
0.75 & 0.40993 & & 0.4017(41) & 1.39 & & & 0.4101(25) & 0.69\\
0.9 & 0.57983 & & 0.5778(36) & 1.74 & & & 0.5777(21) & 0.73\\
1 & 0.65552 & & 0.6523(27) & 1.83 & & & 0.6558(22) & 0.69\\
\bottomrule
\end{tabular}
\caption{Ladders at $L=4$, started from uniform configurations at
$\beta=0$. At each coupling, the network trained on the chain of the
previous stage generates every proposal from the uniform terminal with 64
levels, and the chain has 20{,}000 iterations. ``acc.''\ is the path
acceptance, and ``generator + local'' adds one local sweep with the
learned score to every iteration. Errors are block errors of one chain,
and the exact values are from the character expansion.}
\label{tab:ladder-L4}
\end{table}

With the uniform terminal every accepted configuration comes from the
generator, so chains with the generator alone test the path correction of
eq.~\eqref{eq:path-accept} directly. With the networks of the $\beta=0.3$
and $\beta=1$ stages of the ladder with the generator alone, six chains of
250{,}000 iterations at
each coupling give $\langle E\rangle=-0.31931(20)$ and
$\langle\Upsilon\rangle=0.01682(17)$ at $\beta=0.3$, and
$\langle E\rangle=-1.41134(28)$ and $\langle\Upsilon\rangle=0.65537(29)$ at
$\beta=1$. All four agree with the exact values of
table~\ref{tab:ladder-L4} within 1.4 standard errors, and at $\beta=1$ the
errors are 0.02\% of the energy density and 0.04\% of the helicity
modulus. At every coupling the path acceptances of the three ladders with
the generator alone agree within 0.011.

\subsection{Ladders at \texorpdfstring{$L=8$}{L=8}}
\label{sec:results-L8}
The ladders at $L=8$ start again from uniform configurations at $\beta=0$,
independently of the ladders at $L=4$. The relative effective sample size
of the first step, $\beta=0\to0.15$, is 0.195, close to $0.68^4=0.21$,
the fourth power of its value at $L=4$, as expected when
$\log(N_w/N_k)$ is proportional to the volume. We ran three ladders: the generator
alone, the generator with the local sweep, and the generator alone with
the larger network.

\begin{figure}[t]
\centering\includegraphics[width=\textwidth]{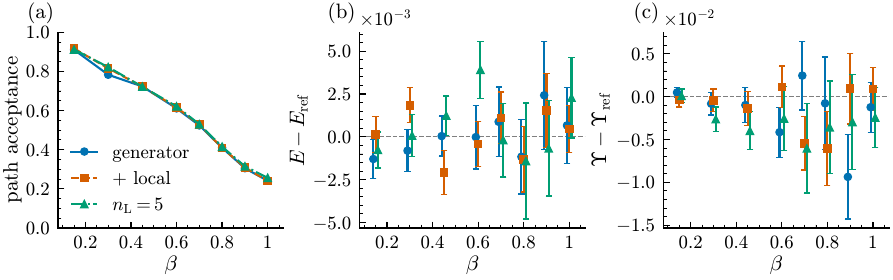}
\caption{Ladders at $L=8$, started from uniform configurations at
$\beta=0$. (a) Path acceptance of the generator, which proposes every
configuration from the uniform terminal with 256 levels. (b), (c)
Deviation of the energy density and of the helicity modulus from the
reference values of the heat-bath and overrelaxation calculation of
appendix~\ref{app:reference}. The legend of (a) applies to all panels:
``generator'' is the ladder with the generator alone, ``+\,local'' adds
one local sweep with the learned score to every iteration, and
$n_{\rm L}=5$ denotes the ladder with the generator alone and the larger
network with five smearing layers and width 64. Each point is one chain
of 20{,}000 iterations with block errors. Points are shifted horizontally
for visibility.}
\label{fig:ladder-L8}
\end{figure}

All three ladders reproduce the reference values at all eight couplings
(figure~\ref{fig:ladder-L8}). For eight degrees of freedom, $\chi^2$ for
the energy and the helicity modulus is $2.9$ and $7.5$ with the generator
alone, $7.6$ and $6.2$ with the local sweep, and $8.4$ and $9.7$ with the
larger network; the largest deviation is 2.4 standard errors. The path
acceptance at $\beta=1$ is 0.24, and 0.26 with the larger network. At
$\beta=1$ the generator alone gives $\tauint(E)=3.7$ and
$\tauint(\Upsilon)=2.4$, longer than one sweep of the control at $L=8$
for $E$ and shorter for $\Upsilon$, and with the local sweep $1.3$ and
$1.3$, about equal to two sweeps of the control for $E$ and shorter for
$\Upsilon$ (table~\ref{tab:production}).

To judge the quality of the self-learned generator, we compare the
networks of the last stage of the two ladders with $n_{\rm L}=3$ with the
network trained at $L=8$ (section~\ref{sec:experiments}). With 256 levels
from the uniform terminal they give path acceptances of 0.241 and 0.240,
against 0.239 for the network trained at $L=8$. We also evaluate the
networks on identical noisy configurations. Since the conditional mean of
the training target is $s^\star_{\beta,\sigma}$, the difference of the
mean squared deviations of two networks from the target estimates the
difference of their squared score errors. For $\sigma\geq0.12$ this
difference is below $3\times10^{-4}$ per component and mostly within two
standard errors. At $\sigma=0.05$ the ladder networks are worse by up to
$1.5\times10^{-5}$, about a fifth of the mean squared deviation of the
network trained at $L=8$; longer training lowers this deviation at
$\sigma=0.05$ (section~\ref{sec:results-acceptance}). We conclude that, starting from
$\beta=0$ and the action, the ladder reaches a generator of the same
quality as one trained on a chain at the target coupling.

\subsection{Larger lattices at \texorpdfstring{$\beta=1$}{beta=1}}
\label{sec:results-production}
Table~\ref{tab:production} compares the production chains, which use the
network of the $L=4$ ladder (section~\ref{sec:experiments}) without
retraining, with the control and the
reference values on lattices up to $L=16$. All entries, including those
of the control, agree with the reference values within 2.2 standard
errors. We find that with $V/N\approx1$ the path acceptance stays between
0.41 and 0.45 from $L=8$ to $L=16$, although the network has been trained
only at $L=4$. The networks of the last stages of the three ladders at
$L=4$ with the generator alone (section~\ref{sec:results-L4}) give path
acceptances of 0.451, 0.447 and 0.458 at $L=8$ and 0.413, 0.413 and 0.412
at $L=16$. An iteration of the generator with the local sweep consists of a path
proposal and a sweep of the local kernel with the learned score. With the
same number of updates per iteration, two sweeps of the control give
shorter autocorrelation times than this iteration, for $E$ at every
volume and for $\Upsilon$ at $L=8$ and $12$; at $L=16$, chains of
20{,}000 iterations do not determine $\tauint(\Upsilon)$. A path proposal
alone, the generator without the local sweep, gives autocorrelation times
of $E$ and $\Upsilon$ 1.9 and 1.6 times those of one sweep of the control
at $L=8$, and 1.3 and 0.85 times at $L=12$.

\begin{table}[t]
\centering\small\setlength{\tabcolsep}{5pt}
\begin{tabular}{cl c c ll cc}\toprule
$L$ & kernel & chains & acc. & \multicolumn{1}{c}{$\langle E\rangle$} & \multicolumn{1}{c}{$\langle\Upsilon\rangle$} & $\tauint(E)$ & $\tauint(\Upsilon)$\\\midrule
8 & generator + local & 4 & 0.451 & $-1.35881(82)$ & $0.5734(21)$ & 1.76 & 2.96\\
 & generator & 2 & 0.450 & $-1.3573(57)$ & $0.5716(82)$ & 4.17 & 5.86\\
 & local, action force & 4 & & $-1.35838(99)$ & $0.5722(22)$ & 2.25 & 3.56\\
 & local $\times2$, action force & 4 & & $-1.3588(13)$ & $0.5741(26)$ & 1.29 & 1.93\\
 & reference & & & $-1.35759(15)$ & $0.57153(26)$ & & \\\midrule
12 & generator + local & 4 & 0.411 & $-1.34214(86)$ & $0.5238(31)$ & 3.52 & 7.94\\
 & generator & 2 & 0.415 & $-1.3414(33)$ & $0.5204(41)$ & 6.45 & 12.5\\
 & local, action force & 4 & & $-1.3423(11)$ & $0.5154(49)$ & 4.82 & 14.8\\
 & local $\times2$, action force & 4 & & $-1.34295(96)$ & $0.5259(21)$ & 2.44 & 5.25\\
 & reference & & & $-1.34266(12)$ & $0.52592(39)$ & & \\\midrule
16 & generator + local & 4 & 0.413 & $-1.3362(13)$ & $0.4872(70)$ & 5.04 & 20.3\\
 & local, action force & 4 & & $-1.33714(75)$ & $0.4990(28)$ & 6.46 & 17.7\\
 & local $\times2$, action force & 4 & & $-1.33664(84)$ & $0.4941(83)$ & 3.49 & 12.8\\
 & reference & & & $-1.33598(5)$ & $0.49279(26)$ & & \\
\bottomrule\end{tabular}
\caption{Production chains at $\beta=1$ with the last-stage network of the
$L=4$ ladder run with the generator alone, used without retraining. Path
proposals use the
shared terminal with $\sigma_N=0.8$ and $N=64$, $128$ and $256$ levels at
$L=8$, $12$ and $16$. ``local, action force'' is the local kernel of
section~\ref{sec:generation-step} with the drift $-\nabla S_\beta$ and no
network, with one sweep per iteration, and ``local $\times2$'' applies two
such sweeps per iteration. Each chain has 20{,}000 iterations. Errors are from the spread
between chains, with one degree of freedom in the rows with two chains,
and $\tauint$ is the mean over chains. Single-chain values of
$\tauint(E)$ lie within about 20\% of the mean, and those of
$\tauint(\Upsilon)$ within a factor of about 1.5 at $L=8$ and $12$; at $L=16$,
$\tauint(\Upsilon)$ of single chains ranges from 9 to 45 for generator +
local and from 14 to 23 for one sweep of the control. Reference values are from the
heat-bath and overrelaxation calculation of appendix~\ref{app:reference}.}
\label{tab:production}
\end{table}

A network trained at a larger volume has a smaller error at noise levels
above $\sigma=0.8$ (section~\ref{sec:results-acceptance}) and allows
larger moves. With the network trained at $L=8$, the shared terminal at $\sigma_N=3.2$ and $N=256$
levels give at $L=8$, in four chains of 20{,}000 iterations, a path
acceptance of 0.244(1) and, with the local sweep, $\tauint(E)=1.23(1)$,
about equal to that of two sweeps of the control, and
$\tauint(\Upsilon)=1.23(3)$, shorter (table~\ref{tab:production}). The
means agree with the reference values within 1.9 standard errors.

\subsection{Acceptance, number of levels and network error}
\label{sec:results-acceptance}
\begin{figure}[t]
\centering\includegraphics[width=\textwidth]{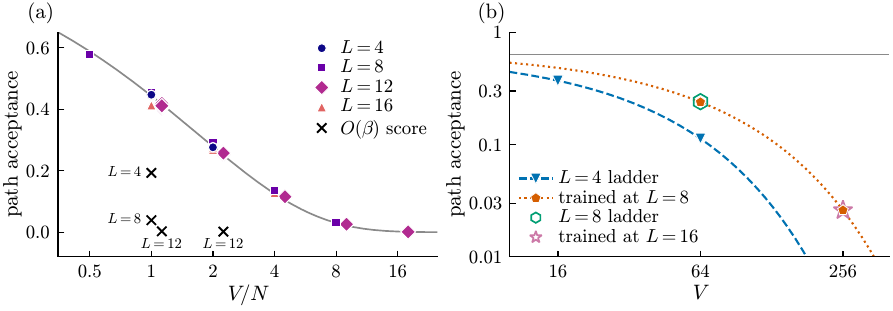}
\caption{Path acceptance at $\beta=1$. (a) Shared terminal at
$\sigma_N=0.8$, for networks trained only at $L=4$ (the $n_{\rm L}=3$
network of table~\ref{tab:network-transfer} and networks of the same
architecture trained by ladders at $L=4$), against $V/N$. Symbols give
the lattice size $L$ of the chain, and points of
different networks and chains at the same $L$ and $V/N$ overlap. Crosses
use the $O(\beta)$ score $e^{-\sigma^2}(-\nabla S_\beta)$ without
training, at $L=4$, $8$ and $12$ with $V/N\approx1$ and at $L=12$ with
$V/N=2.25$. The curve is eq.~\eqref{eq:acceptance-model} with $c=2.35$ and
$b=0$. (b) Uniform terminal with $V/N=1/4$, against the volume, for the
network of the $L=4$ ladder, the network of one $L=8$ ladder with
$n_{\rm L}=3$, and the networks trained at $L=8$ and $16$. The dashed and
dotted curves are eq.~\eqref{eq:acceptance-model} with $c$ and $b$
determined from the two filled points of the network of the $L=4$ ladder
and of the network trained at $L=8$, respectively. The open symbols lie on
the filled symbols of the network trained at $L=8$ and are not used for
the curves. The horizontal line is $b=0$ with $c=3.6$, the mean of the two
values of $c$ weighted by their errors. Binomial errors of the acceptance
are smaller than the symbols.}
\label{fig:acceptance}
\end{figure}

With the shared terminal at $\sigma_N=0.8$, the path acceptance of
networks trained only at $L=4$ lies close to one curve as a function of $V/N$
for lattices from $L=4$ to $16$ and $N$ from 8 to 256
(figure~\ref{fig:acceptance}(a)). The curve is
eq.~\eqref{eq:acceptance-model} with $b=0$ and $c=2.35(1)$, fitted to
these points. Points with the same $L$ and $N$ agree within their binomial
errors, but the points deviate from the curve by up to 0.03, more than
these errors ($\chi^2/\mathrm{dof}=14$), and the error of $c$ is scaled by
$\sqrt{\chi^2/\mathrm{dof}}$. With $b$ as a second parameter, and its error scaled in the same
way, the fit gives $b=0.0009(2)$. The
smoothed score to first order in $\beta$,
$e^{-\sigma^2}(-\nabla S_\beta)$, which we call the $O(\beta)$ score,
behaves differently. On $L=4$, $8$ and $12$ with $V/N\approx1$ its
acceptance falls from 0.19 at $L=4$ to 0.04 at $L=8$ and 0.003 at
$L=12$, which corresponds to $b$ between 0.23 and 0.28. For $\sigma\leq0.8$
the error of the trained network per site is thus two orders of magnitude
below that of the $O(\beta)$ score and small compared with the
discretization term. The networks trained
at $L=8$ and $16$ raise the acceptance at $L=16$ with $N=256$ only from
0.413 to 0.432 and 0.441, close to the 0.443 of the curve with $b=0$.

Proposals from the uniform terminal use noise levels up to $\sigma=3.2$,
where the network error is visible (figure~\ref{fig:acceptance}(b)). At
fixed $V/N=1/4$ the acceptance falls with the volume. For two networks we
determine $c$ and $b$ from the acceptances on two lattices, $L=4$ and $8$
for the network of the $L=4$ ladder and $L=8$ and $16$ for the network
trained at $L=8$, and propagate the binomial errors of the acceptances.
Two acceptances fix the two parameters, so the test of
eq.~\eqref{eq:acceptance-model} here is whether the values of $c$ for the
two networks agree; the shared terminal above tests the dependence on
$V/N$ with many points. We
obtain $c=3.7(3)$ and $b=0.143(3)$ for the network of the $L=4$ ladder, and
$c=3.1(7)$ and $b=0.074(2)$ for the network trained at $L=8$. The two values
of $c$ agree within their errors, as they should if $c$ depends only on the
levels, and their mean weighted by the errors, $3.6(3)$, exceeds the value
2.35 of the shared terminal because the levels of the uniform terminal
span a wider range of $\sigma$. Training at $L=8$ instead of $L=4$ halves
$b$ and doubles the acceptance at $L=8$, from 0.11 to 0.24. With the $c$
of the network trained at $L=8$, the networks of the two ladders at $L=8$
with $n_{\rm L}=3$, trained from $\beta=0$, both have $b=0.074$, equal
within the binomial errors to that of the network trained at $L=8$, and
the larger network of the third ladder has $b=0.068$. At $L=16$ with $N=1024$ the networks
trained at $L=8$ and $16$ have the same acceptance, 0.026, and hence the
same $b$, so training at $L=8$ already reaches the $b$ of training at
$L=16$. Three times more training steps, alone or with four times more
configurations, lower the mean squared deviation of the network trained
at $L=8$ at $\sigma=0.05$ from $0.7\times10^{-4}$ to $0.3\times10^{-4}$ per
component and change it at $\sigma\geq0.6$ by less than $4\times10^{-4}$.
The reduction of $b$ by training at $L=8$ comes from noise levels between
$\sigma=1.1$ and $2$, where the score error per component decreases by up
to $0.03$.

\section{Discussion}
\label{sec:discussion}
The ladder relies on the overlap of the distributions at neighbouring
couplings. Action differences are extensive, so $\log(N_w/N_k)$ is roughly
proportional to the volume, as the first step of the $L=8$ ladder shows
(section~\ref{sec:results-L8}), and the coupling increment must shrink as
the volume grows~\cite{Ferrenberg:1988reweighting,Iwami:2015multipoint,Weigel:2021ckd}.
Ref.~\cite{Wang:2026pmv} reports, for a data-free sampler that reweights
by path weights, a relative effective sample size of order $10^{-3}$ in
the broken phase of two-dimensional scalar $\phi^4$ theory at $L=64$. In
\SLD the replay is refreshed by an MH chain at every stage, so the weights
act across one step only, and the step can be chosen from $N_w/N_k$ itself,
as at $L=8$ from $\beta=0.6$ on. A second route to large volumes is to run
the ladder on a small lattice and generate on the large one: with the
shared terminal at $\sigma_N=0.8$ and $V/N=1$, the network of the $L=4$ ladder reaches
at $L=16$ nearly the path acceptance of the network trained at $L=16$
(section~\ref{sec:results-acceptance}).

The learned score and the action force play different roles in the two
kernels. The MH ratio of the local kernel contains the action, and the
action force is the limit of the score as $\sigma\to0$; the network is
trained to represent the score of the noise-smoothed distribution, which
at finite $\sigma$ differs from this force.
Reverse diffusion, in contrast, needs the smoothed score at every noise
level, and the action does not give it in closed form. To first order in
$\beta$ it is the $O(\beta)$ score $e^{-\sigma^2}(-\nabla S_\beta)$, whose
error per site for $\sigma\leq0.8$ is two orders of magnitude larger than
that of the trained network (section~\ref{sec:results-acceptance}). The
training target of eq.~\eqref{eq:tsm-blend} uses the action as well, but
its weight on the action force is small at large noise,
$\lambda_\sigma=0.047$ at $\beta=1$ and $\sigma=3.2$. The network error
that lowers the acceptance of the uniform terminal comes from noise levels
above $\sigma=0.8$. Objectives that use the
action more effectively at large $\sigma$ may reduce this error. In the XY
model the action force is cheap. A learned generator matters more where the
force is expensive, as in first-principles molecular
dynamics~\cite{Nagai:2020slhmc} and in non-Abelian gauge theory with
dynamical fermions~\cite{Nagai:2020jar}, for which self-learning updates
were developed.

Autocorrelation depends on the size of the accepted moves as well as on the
acceptance~\cite{Roberts:1996mala,Roberts:1998mala,Pasarica:2010esjd}. With
the uniform terminal every proposal is independent of the current
configuration, and autocorrelation comes only from rejections
(section~\ref{sec:results-L4}). The shared terminal at $\sigma_N=0.8$ with
$V/N\approx1$ keeps the acceptance above 0.4 up to $L=16$, but the proposal
retains much of the current configuration, and $\tauint(\Upsilon)$ grows
with the volume. The network trained at $L=8$ allows $\sigma_N=3.2$ with
$N=256$ at $L=8$, where the proposal nearly forgets the current
configuration; with the local sweep, $\tauint(\Upsilon)$ is then shorter than with two
sweeps of the local kernel with the action force, whereas at
$\sigma_N=0.8$, with the network of the $L=4$ ladder and $V/N\approx1$,
it is longer (section~\ref{sec:results-production}).
The terminal noise thus trades the size of the move against the acceptance, and
the network error sets how large $\sigma_N$ can be for a given number of
levels.

The chains of the ladder are Markov chains as well, and slow modes that
they explore poorly also enter the replay. Since each chain is exact for
any network, such a deficit of the training data affects the acceptance
and the autocorrelation times rather than the sampled distribution.
Our results are limited to $\beta\leq1$ and $L\leq16$, on the
high-temperature side of the BKT transition; critical slowing down near the
transition has not been studied. Higher-order reverse steps, the
counterpart of higher-order integrators in molecular dynamics, may reduce
the term $cV/N$ of eq.~\eqref{eq:acceptance-model} and slow the growth of
$N$ with $V$, and noise schedules
designed by controlling the total variance and the signal-to-noise
ratio~\cite{Kahouli:2025tvsnr} can be judged by their effect on the path
acceptance.

\section{Conclusion}
\label{sec:conclusion}
We presented the self-learning diffusion sampler \SLD, in which a diffusion
model is trained along a ladder of couplings from exact samples at
$\beta=0$ on the MH chain that it generates, without samples or a
variational objective at the target coupling. Its MH test acts on the extended space of the
configuration and its noising path, which makes $p_\beta$ the invariant
distribution of the chain for any network. In the XY model, ladders from $\beta=0$ to
$1$ at $L=4$ and $8$ reproduce exact and reference values at every coupling,
and the generator of the $L=8$ ladder matches the acceptance of a network
trained on an exact chain at the target coupling. The path acceptance
follows the relation known from HMC, with terms set by the number of noise
levels and by the network error per site, and with the shared terminal at
$\sigma_N=0.8$ and $N\propto V$ a network trained at $L=4$ generates at
$L=16$ with an acceptance of 0.41. Ladders at $L=4$ repeated with different seeds change the path acceptance by at
most 0.011 at every coupling, so that the self-learning from $\beta=0$ is
reproducible.

Two directions are natural next steps. First, higher-order reverse
steps and objectives that use the action at large $\sigma$ may reduce the two
terms of the acceptance relation. Second, gauge theories, where
gauge-covariant networks are
available~\cite{Nagai:2021bhh,Favoni:2020reg,Nagai:2025rok,Tomiya:2025quf}
and the force can be expensive, would test the method where no cluster
algorithm exists.

\section*{Data and code availability}
The numerical data supporting the findings of this study are available from
the author upon reasonable request. The simulation, analysis, and plotting
codes will be made public after the publication of this paper.
\section*{Acknowledgments}
This work was supported by JSPS KAKENHI Grants No.~20K14479, No.~22H05111,
No.~22K03539, No.~22H05112, and JST BOOST Grant No.~JPMJBY24F1.
This work was supported by JST, ARiSE, Japan Grant Number JPMJAR2633.
OpenAI's ChatGPT and Codex and Anthropic's Claude, including Claude Opus 5.5,
were used, under the author's direction, for writing the simulation
and analysis code, the numerical calculations, the figures, the literature
survey, and the preparation of the manuscript. The author assumes
responsibility for the scientific claims, algorithms, numerical results,
citations, and final content of the manuscript.
\appendix
\section{Ergodicity of the kernels and truncation of the image sums}
\label{app:exactness}
The path proposal of section~\ref{sec:path-mh} and every level of the local
kernel of section~\ref{sec:generation-step} are MH steps that satisfy
detailed balance with respect to $p_\beta$. The path proposal, the local
sweep and their composition therefore leave $p_\beta$ invariant, although
the sweep need not satisfy detailed balance itself. Convergence to
$p_\beta$ from any initial configuration requires in addition
irreducibility and aperiodicity~\cite{Roberts:1996mala,Sokal:1997}. At
finite volume the configuration space is compact and $p_\beta$ is
positive. The proposal densities, eqs.~\eqref{eq:forward-kernel},
\eqref{eq:reverse-kernel} and~\eqref{eq:q}, are built from wrapped
Gaussians with finite drift and positive variance, so that every proposal
has a positive density between any two configurations and a positive MH
acceptance probability. Each kernel is therefore irreducible and
aperiodic.

The sum over images in eq.~\eqref{eq:wrapped} is evaluated after $u$ is
reduced to its principal interval, and is truncated at
$|\nu|\leq\nu_{\max}$. For $|u|\leq\pi$ the omitted images lie at distances
of at least $(2\nu_{\max}+1)\pi$. Bounding them by a geometric series
gives, for the omitted part $R_v(u)$,
\begin{align}
 \frac{R_v(u)}{G_v(u)}\leq
 \frac{2\exp[-2\nu_{\max}(\nu_{\max}+1)\pi^2/v]}{1-\exp[-4(\nu_{\max}+1)\pi^2/v]},
 \label{eq:tail-bound}
\end{align}
where the principal image $\nu=0$ bounds $G_v$ from below. The local kernel
uses $\nu_{\max}=3$ with $v=2h_i\leq1.28$, which gives a bound below
$10^{-79}$ per component. The path proposals and the training of
section~\ref{sec:experiments} use $\nu_{\max}=5$ with $v\leq10.24$, which
gives a bound below $10^{-24}$. Densities and MH ratios are evaluated in
logarithmic form in Float64.

\section{Reference values}
\label{app:reference}
The reference values in tables~\ref{tab:ladder-L4}
and~\ref{tab:production} and in figure~\ref{fig:ladder-L8}
were computed with codes written independently of the samplers of this
paper.

\paragraph{Character expansion.}
Inserting
$e^{\beta\cos\Delta_\mu\phi_x}=\sum_{n_{x,\mu}\in\mathbb Z}I_{n_{x,\mu}}(\beta)e^{in_{x,\mu}\Delta_\mu\phi_x}$
for every link $(x,\mu)$ and integrating over the angles gives
$Z=\sum_{\{n\}}\prod_{x,\mu}I_{n_{x,\mu}}(\beta)$, where the integer link
currents $n_{x,\mu}$ are conserved at every site. We take the horizontal
currents crossing a column cut as the state; current conservation then
fixes the vertical currents of a column up to a constant, which is summed
over. This defines a transfer matrix $\mathcal T$ that is block diagonal
in the total current $Q$, and $Z=\sum_QZ_Q$ with
$Z_Q=\operatorname{Tr}\mathcal T_Q^L$. We truncate the horizontal currents
at $|n_{x,1}|\leq n_{\max}$. At $L=4$ the energy changes by less than
$10^{-12}$ between $n_{\max}=4$ and $5$, and at $L=6$ by less than
$10^{-12}$ between the two largest truncations. The energy density follows
from $\langle E\rangle=-V^{-1}\partial\ln Z/\partial\beta$ by finite
differences. A twist of the links crossing one column cut by an angle
$\vartheta$ multiplies each current sector by $e^{-iQ\vartheta}$, which
gives for the helicity modulus of eq.~\eqref{eq:helicity}
\begin{align}
 \langle\Upsilon\rangle=\frac{\langle Q^2\rangle}{\beta},\qquad
 \langle Q^2\rangle=\frac{\sum_QQ^2Z_Q}{Z}.
\end{align}
On $2\times2$ and $3\times3$ lattices the results agree with direct
numerical integration to $10^{-14}$, and at $L=4$ and $\beta=0.5$ with a
transfer matrix in the angle basis to $3\times10^{-10}$ for the energy and
$2\times10^{-9}$ for the helicity modulus.

\paragraph{Heat-bath Monte Carlo.}
We use one single-site heat-bath sweep with von Mises sampling followed by
one microcanonical overrelaxation sweep. For the couplings of the $L=8$
ladder in figure~\ref{fig:ladder-L8} and for $\beta=1$ at $L=8$, $12$ and
$16$ in table~\ref{tab:production}, we use four chains of $5\times10^5$
sweeps at each coupling ($2\times10^6$ at $L=16$) after $10^5$
thermalization sweeps. Errors are obtained from 50 blocks per chain and
agree with the spread between chains. The same update, with 20 chains of
$2\times10^6$ to $4\times10^6$ sweeps after $10^5$ thermalization sweeps,
agrees with the exact energy and helicity modulus at $L=4$ for
$\beta=0.30$, $0.35$, $0.40$, $0.45$ and $0.50$ and at $L=6$ for $\beta=0.5$, with $\chi^2=9.9$ for
12 comparisons. As a check of the normalization of $\Upsilon$, the same
calculation at $L=4$ and $\beta=0.15$, $0.6$ and $1$ agrees with the exact
values within two standard errors.

\clearpage
\providecommand{\href}[2]{#2}\begingroup\raggedright\endgroup

\end{document}